\documentstyle[aps]{revtex}

\def\U#1{\mathop{U}\limits^{(#1)}}  
\def\D#1{\mathop{D}\limits^{(#1)}}

\def\P{{\cal P}}

\def\D{{\cal D}}  
  
\def\12{{1\over 2}}

\def\po{p_\theta}

\title{ The BRST-BFV  method for non-stationary systems } 
 
\author{ J. Antonio Garc\'ia and  J. David Vergara} 
\address{ Instituto de Ciencias Nucleares \\
Universidad Nacional Aut\'onoma de M\'exico \\
 Apartado Postal 70-543, 04510 
 M\'exico, D.F.} 
\author{and}
\author{Luis F. Urrutia \thanks{On sabbatical leave from Instituto de Ciencias Nucleares, Universidad Nacional Aut\'onoma de M\'exico,
Circuito Exterior, C.U., 04510 M\'exico, D.F.}}
\address{Departamento de F\'\i sica \\
Universidad Aut\'onoma Metropolitana-Iztapalapa \\
          Apartado Postal 55-534, 09340 M\'exico, D.F. \\
\
 and  \\
Facultad de F\'\i sica  \\
Universidad Cat\'olica de Chile \\
 Casilla 306, Santiago 22 , Chile}

\begin{document} 
\maketitle 
\begin{abstract} 
Starting from an  associated reparametrization-invariant action,   the generalization of the BRST-BFV method for the case of non-stationary systems  is constructed. The extension of the  Batalin-Tyutin conversional aproach  is also considered in the non-stationary case. In order to illustrate these ideas, the propagator for  
the time-dependent  two-dimensional rotor is calculated  by reformulating the problem as a system with only first-class constraints and subsequently using the BRST-BFV prescription previously obtained.
 \end{abstract}
  
\section{Introduction} 

The extension of the Dirac method to include  constraints that 
depend explicitly on time was developped in \cite{mu,gi,ev}, but a systematic construction of the BRST-BFV quantization procedure \cite{brst}  together with the Batalin-Tyutin conversional approach \cite{ba} has not been given in detail yet \cite{baly}. We start  the discussion of these problems  with  
a brief introduction to    the standard Dirac procedure for  systems with time-dependent   constraints.

Let us consider a system  described by a $2n$-dimensional phase space $ (q^i, p_j)$,  a canonical Hamiltonian $ H_0 (q,p,t)$,  a set of first-class constraints $ \left\{ \psi_A(q,p,t) \right\}$,  and  a set of second-class constraints  $\{\phi_\alpha(q,p,t)\}$ . We  assume that  all these functions in phase space are explicitly time-dependent. The canonical description of such system is given by the action
\begin{equation}
S=\int ( p_i {\dot q}^i- H_0-\lambda^\alpha\phi_\alpha - \mu^A\psi_A)dt,
\end{equation}
where $ \lambda^\alpha$  and $ \mu^A$ are Lagrange multipliers. We also assume that   the consistency conditions
\begin{equation}
{d{\phi_\alpha} \over dt}\approx0, \ \ \ \ {d{\psi_A} \over dt}\approx0,
\end{equation}
 are identically satisfied in order to guarantee  that the Dirac procedure has been completed and no further constraints arise. The time evolution of an arbitrary time-dependent function $ F(q,p,t)$ is given by
\begin{equation}
{d{F} \over dt}={\partial {F}\over \partial t} +\left\{ F, \  H_T \right\},
\end{equation}
where  $ H_T=H_0 + \mu^A\psi_A +\lambda^\alpha\phi_\alpha $ is the total Hamiltonian. The conservation in time of the second-class constraints allow us to solve for the corresponding Lagrange multipliers leading to
\begin{equation}
{d{F} \over dt}={\partial {F}\over \partial t} +\left\{ F, \  H_0 \right\}^*+\mu^A\left\{F, \ \psi_A \right\}^*-\{F,\phi_\alpha\}C^{\alpha\beta}\frac{\partial \phi_\beta}{\partial t},
\end{equation}
where $\{ \, \ \}^*$ denotes de Dirac bracket in the $ (q^i,p_j)$ phase space 
\begin{equation}
\{ \ , \ \}^*= \{ \ , \ \} - \{ \ , \phi_\alpha\}C^{\alpha\beta}\{\phi_\beta,  
\ \},
\end{equation}
with  the matrix $C_{\alpha\beta}= \left\{ \phi_\alpha, \ \phi_\beta \right\}$ been  time-dependent  in general. The matrix $ C^{\alpha\beta}$ is the inverse of $ C_{\alpha\beta}$.  
The last term in the right-hand side of Eq.(4) breaks the canonical structure of the evolution equation {\it i.e.} the time 
evolution is not  any more the unfolding of a canonical transformation. Moreover, it can be readily shown that is not possible at this stage to absorb this extra term inside the Dirac bracket by adding a suitable  piece  to $ H_0$ \cite{mu}.

 There are at least two ways of recovering the canonical structure of the evolution equations in the non-stationary  case. 
One approach is due to Mukunda \cite{mu}  and  reduces to finding an apropriate change  of 
coordinates in phase space.  The basic idea  is to include all second-class constraints as new coordinates. This initial set  is subsequently completed by adding  extra  variables  in such a way to end up with a set of  independent invertible  functions that can be considered as new coordinates (non-necessarily canonical)  for the problem. The additional coordinates turn out to be the true degrees of freedom of the system provided there are no first-class constraints present. In the new coordinate patch it is possible to redefine the Hamiltonian in such a way that the equations of motion are given by the corresponding Dirac bracket only. A drawback of  this method is that the adecuate  non-canonical transformation is in general difficult to perform. Also,  the identification of the true degrees of freedom might be completely
non-trivial, or even unconvenient  as it is    in the case where some important symmetries turn out to be non-manifest after such identification is made.  

Another approach to the same problem is given by Gitman and Tyutin \cite{gi} and consists in  adding a couple of new canonically conjugated  variables which are the time  $t$ together with its momentum ${\pi_t}$ in such a a way that  $\{t,\pi_t \}=1$. Then,   the evolution equation (4) can be   written as a single Dirac bracket
\begin{equation}
\frac{dF}{dt} \approx \{F,H_0+\pi_t +\mu^A \psi_A\}^*. 
\end{equation}

 A conceptual  point which must be addressed  in  this proposal is the question of the  equivalence between the original theory and the extended one which, in principle,  has two more degrees of freedom in phase space.

In this paper we clarify this equivalence  by showing that the new  variables $ t$ and $ \pi_t$  arise quite naturally by  rewriting a general time-dependent system in parametrized form, as proposed in Ref.\cite{ht}.  In this way,  $t$ is now an extra coordinate that evolves according to some parameter $s$ in a canonical fashion. Such parametrized description involves an additional  first-class constraint associated with the reparametrization invariance, which is just $H_0 +\pi_t$. It is precisely this constraint which 
supresses the additional degrees of freedom, thus getting us back to the original phase space.
Following this idea, we first  extend the system into an equivalent one which behaves canonically
with respect to the new evolution parameter $s$. For this system we are able to construct the BRST-BFV  propagator in the standard fashion and subsequently we proceed to integrate over the additional
variables introduced, in such a way to cast the propagator in terms of the initial variables  only, which  include the explicit time dependence of the hamiltonian and the constraints. This method provides a natural way to  construct  both the BRST charge and the BRST Hamiltonian in the non-stationary case and also allows  to implement the BFV prescription in this situation.

Another important question to be addressed is whether or not  the FV theorem is still applicable in the non-stationary case. Our final result for the BRST-BFV propagator still contains an arbitrary fermionic gauge which is the remaining of the  partial choice of the fermionic gauge in the extended canonical system, which has been fixed only in the sector of  the additional  variables that we want to eliminate.  This final expression is subsequently  used to study   the validity of the  FV theorem  in the non-stationary case. 

In this work we do not consider the quantization of systems with second class constraints through the Dirac bracket approach. Instead, we take the point of view of promoting the BRST-BFV  quantization procedure to the status of a universal  prescription for quantizing any constrained system.  For this reason, we  are also interested in  extending   the conversional approach of Batalin and Tyutin \cite{ba} to the  non-stationary case  and  we also provide the necessary prescriptions. 

The paper is organized as follows: in section 2  we show  how   the proposal of Gitman and Tyutin [2] arises naturally from a parametrized formulation of the original non-stationary system. 
In section 3 we use this method of  dealing with time-dependent systems to show that   the  construction of the extended Hamiltonian in  the conversional approach can be performed by starting from  the corresponding  expressions already known for the time-independent case; except that  now the boundary condition in the recursive expressions for the Hamiltonian  are changed from $H_0$ to $H_0+\pi_t$. Section 4 contains the main result of the paper which is the construction of the BRST-BFV formalism  for
non-stationary systems, starting from the canonical prescription for the equivalent parametrized system.  Finally, in section 5 we consider the simple example of a two-dimensional  rotor with time-dependent
radius.  We use the conversional approach to reformulate the problem as a system with  first-class constrains only   and we  apply the BRST-BFV quantization  method developped in the previous section.  We calculate the quantum mechanical propagator, obtaining the well-known result.

\section{ Reformulation of the problem in terms of a  parametrized  action}

In this section we  consider the following   reparametrization-invariant  version of the action  (1)
\begin{equation}
S=\int \Bigg( p_i {dq^i \over ds}- {dt \over ds }(H_0(q,p,t) + \lambda^\alpha\phi_\alpha + \mu^A\psi_A)\Bigg)ds,
\end{equation}
where  we demand that $ {d{t} \over ds}\neq0$. In this approach we are promoting $t$ to the status of a coordinate, with its corresponding canonical momentum $\pi_t$, so that the new phase space is  of dimensions $2n+2$. 

 Applying  the standard  Dirac procedure to the action (7)  
we obtain  the  primary constraints
\begin{equation}
\psi=\pi_t+H_0(q,p,t)\approx 0, \ \ \  \phi_\alpha(q,p,t)\approx 0, \ \ \  \psi_A(q,p,t)\approx 0.
\end{equation}
The corresponding canonical Hamiltonian in zero and  the total Hamiltonian ${\tilde H}_T=\tilde{\mu}(\pi_t+ H_0) +\tilde{ \mu}^A\psi_A +\tilde{\lambda}^\alpha\phi_\alpha $  describes the evolution with respect to the parameter $s$.  Let us  calculate the  evolution of the above primary constraints, observing that none of them has an explicit $s$ dependence. Our first result is 
\begin{eqnarray}
\nonumber {d{\psi_A} \over ds} & \approx & \tilde\mu\left\{ \psi_A, \  \pi_t+ H_0\right\}_{p,q,t,\pi_t}=\tilde\mu\left(
{\partial {\psi_A}\over \partial t}+\left\{\psi_A, \ H_0  \right\}_{p,q,t,\pi_t}\right)\\
&=&\tilde\mu\left({\partial {\psi_A}\over \partial t}+\left\{\psi_A, \ H_0  \right\}\right)\approx \tilde\mu{d{\psi_A} \over dt,}
\end{eqnarray}
where the Poisson bracket without subindices is calculated   with respect to the original variables $q^i,p_j$. The second   equality  in Eq.(9) follows  because the involved functions do not depend on $\pi_t$. We conclude  that $ {d{\psi_A} \over ds}\approx0$ identically, in virtue of Eqs.(2). 
Next  we calculate  
\begin{equation}
{d{\phi_\alpha} \over ds}\approx\tilde\mu\left\{ \phi_\alpha, \ \pi_t+H_0 \right\}_{q,p,t,\pi_t} +{\tilde\lambda}^\beta\left\{\phi_\alpha, \ \phi_\beta  \right\}\approx0,
\end{equation}
which allows us  to determine the Lagrange multipliers associated to the second-class constraints
\begin{equation}
\tilde\lambda^\alpha\approx-\tilde{\mu}C^{\alpha\beta}\left\{ \phi_\beta, \ \pi_t+H_0 \right\}_{q,p,t,\pi_t}.
\end{equation}
Finally, we are left with
\begin{equation}
{d{\psi} \over ds}\approx \tilde{\lambda}^\alpha  \left\{ \psi, \ \phi_\alpha\right\}_{p,q,t,\pi_t}
={\tilde\lambda}^\alpha\left\{ \pi_t +\ H_0 ,\phi_\alpha \right\}=  {1\over \tilde  \mu}{\tilde\lambda}^\alpha C_{\alpha\beta}{\tilde\lambda}^\beta,
\end{equation}
which  is again   zero in virtue of the antisymmetry of $C_{\alpha\beta}$.  From the above analysis we conclude: (i) there are no secondary constraints,  (ii) the constraints $\psi_A$ and $\phi_\alpha$ retain their first-class and second-class character respectively, as expected and (iii)  the new constraint $\psi$ is first-class  in the corresponding  Dirac brackets . It is precisely the property (iii) which guarantees the equivalence between the parametrized description (7) of the system and the original formulation (1), allowing the number of true degrees of freedom to remain the same.

Now, let us consider the evolution equation of an arbitrary function $ F(q,p,t)$, which does not depend either on $\pi_t$ or explicitly on the parameter $s$. We have
\begin{equation}
\frac{dF}{ds}=\{F, \ \tilde{H}_T\}_{p,q,t,\pi_t}\approx \tilde{\mu}\{F, \  \pi_t + H_0\}^*_{q,p,t,\pi_t} +\tilde{\mu}^A\left\{ F, \ \psi_A \right\}_{q,p,t,\pi_t},
\end{equation}
where the last Poisson bracket  is identical to the one calculated in the original phase space,  due to the independence of both functions upon $\pi_t$. In particular we obtain
\begin{equation}
 \frac{dt}{ds}=\tilde \mu \neq0,
\end{equation}
which allows us to calculate directly $ {d{F} \over dt}$ in the parametrized formulation. Comparison with Eq.(6), making   the identification ${ \tilde\mu}^A= \tilde\mu  {\mu}^A$, leads to the result
\begin{equation}
\{F, \  \pi_t + H_0\}^*_{q,p,t,\pi_t}\approx{\partial {F}\over \partial t} +\left\{ F, \  H_0 \right\}^* -\{F,\phi_\alpha\}C^{\alpha\beta}\frac{\partial \phi_\beta}{\partial t},
\end{equation}
which shows how the non-canonical contribution of the RHS is incorporated in the canonical description of the extended parametrizad system having  the additional first-class constraint $ \pi_t +H_0\approx0$. An important consequence of Eq.(15), which will be useful for our further purposes, is the statement 
\begin{equation}
\{F, \  \pi_t + H_0\}_{q,p,t,\pi_t}\approx{\partial {F}\over \partial t} +\left\{ F, \  H_0 \right\}.
\end{equation}
 
\section{The Batalin-Tyutin conversional approach in the non-stationary case}

The conversional  approach \cite{ba,ht} consists in transforming all the original  second-class constraints of a given system into   first-class  constraints in an extended phase space obtained  by adding a proper set of new  canonical variables.  The original Hamiltonian must  also be extended in such a way to  preserve in time  the first-class extended constraints, without generating any new constraint.  For simplicity we assume  that all   degrees of freedom in our theory are  purely bosonic. 

Let us suppose that the system is described by a $2n$-dimensional  phase space with  coordinates $( q^i, p_i)$,  plus  a  Hamiltonian $H_0(q,p,t)$ . We also assume for simplicity that    the theory contains only second-class constraints  $ \phi_\alpha(q,p,t)$, which  are  functionally independent.  Following Batalin and Tyutin \cite{ba}, we introduce  new degrees of freedom
$\Phi^\mu$, in  the same number as the original second-class constraints, 
with the following non-zero  Poisson brackets
\begin{equation}
\{q^i,p_i\}=\delta^i_j \ , \qquad\{\Phi^\mu,\Phi^\nu\}=\omega^{\mu\nu},
\end{equation}
where the constant matrix $\omega^{\mu\nu}$ is antisymmetric and invertible.
Denote by $\tau_\alpha=\tau_\alpha (p,q,\Phi,t)$  the resulting  first class-constraints in the augmented   phase space $ (q,p)\oplus (\Phi)$,
having the desired  property $ \{\tau_\alpha,\tau_\beta \}\approx0,$ together with  the boundary conditions $ \tau_\alpha(p,q,0,t)=\phi_\alpha(p,q,t)\equiv \tau^{(0)}_\alpha(p,q,t). $
The problem of actually finding the constraints $ \tau_\alpha$ is solved constructively, in complete analogy with the time independent case,  by starting from the  series expansion
$
\tau_\alpha=\sum_{n=0}^{\infty}\tau_\alpha^{(n)}, ( \tau_\alpha^{(n)}\sim \Phi^n),
$
in powers of the variables $\Phi^\mu$,  and demanding the first-class condition to each order in $\Phi$.  The explicit expressions that provide the solution to this problem can be found in \cite{ba},  with  the only difference being that most of the expanssion coefficients will become now  functions of time. 

The next step in the conversional approach is to construct the  augmented  Hamiltonian $ H$ and , in general,  the augmented observables of the corresponding theory. The augmented  Hamiltonian  has  the following properties:
\noindent
(i) it  must reduce to  the original Hamiltonian $H_0  $ in the limit $ \Phi^\mu=0$.

\noindent
(ii) it  must preserve in time the first-class time-dependent constraints $ \tau_\alpha$, so that
\begin{equation}
{d{\tau_\alpha} \over dt}= \frac{\partial \tau_\alpha }{\partial t}+\{\tau_\alpha,H\}_{p,q, \Phi}\approx\{\tau_\alpha,H+\pi_t\}_{p,q,\Phi,t,\pi_t}\approx 0,
\end{equation}
where the first weak equality in Eq.(18) is a direct consequence of  Eq. (16) in  the previous section. This is the basic equivalence which will permit us a direct solution of the non-stationary problem in terms of the known solution for the time independent case. In the latter situation (which we denote by a bar over the relevant quantities), let  us suppose that the original second-class constraints $\phi_\alpha(q,p)$ have already been augmented to a first-class set $ \left\{ \bar \tau_\alpha(p,q,\Phi) \right\}$. The equations that determine the extended Hamiltonian  in this case are
\begin{equation}
\left\{ \bar \tau_\alpha, \bar H \right\}_{(p,q,\Phi)}\approx0.
\end{equation}
The solution of Eqs.(19)  is given in Ref.\cite{ba} in terms of a series expansion
\begin{equation}
\bar H=\sum_{n=0}^{\infty}\bar H^{(n)}, \qquad \bar H^{(n)}\sim \Phi^n,
\end{equation}
with the boundary condition
\begin{equation}
\bar H^{(0)}= \bar H_0.
\end{equation}
The explicit solution is given by
\begin{equation}
\bar H^{(n+1)}=-\frac{1}{n+1}\Phi^\mu\omega_{\mu\nu}X^{\nu\rho}\bar G^{(n)}_\rho, \quad n\ge 0,
\end{equation}
where $\omega_{\mu\nu}$ is the inverse of $\omega^{\mu\nu}  $ defined in Eq.(17) and $ X^{\mu\nu}$ is the inverse of $ X_{\mu\nu}$, which is  a solution of $
X_{\alpha\mu} \ \omega^{\mu\nu} \ X_{\beta\nu}=-C_{\alpha\beta}$. Here $C_{\alpha\beta}= \left\{\bar \tau^{(0)}_\alpha, \ \bar \tau^{(0)}_\beta \right\} $. The recurrence expresions for the functions $G^{(n)}_\rho $ are \cite{ba} 
\begin{eqnarray}
\nonumber
\bar G^{(0)}_\alpha&\equiv& \{\bar \tau_\alpha^{(0)}, \bar H^{(0)}\}_{(p,q)},\\
\nonumber \bar G^{(1)}_\alpha\equiv \{\bar{\tau}_\alpha^{(1)}, \bar H^{(0)} \}_{(p,q)}&+&\{\bar\tau_\alpha^{(0)}, \bar H^{(1)}\}_{(p,q)}+\{\bar\tau_\alpha^{(2)}, \bar H^{(1)}\}_{(\Phi)},\\ \nonumber
\bar G^{(n)}_\alpha\equiv \{\bar \tau_{\alpha}^{(n)}, \bar H^{(0)} \}_{(p,q)} + \sum_{m=1}^n \{&&\bar \tau_\alpha^{(n-m)}, \bar H^{(m)}\}_{(p,q)}+\sum_{m=0}^{n-2}\{\bar \tau_\alpha^{(n-m)}, \bar H^{(m+2)}\}_{(\Phi)}\\   &+&  \{\bar \tau_\alpha^{(n+1)}, \bar H^{(1)}\}_{(\Phi)},
\end{eqnarray}
where  the above   Poisson brackets are calculated with respect to the variables indicated in the corresponding subindices.

Now we go back to the non-stationary problem, where the corresponding condition for the extended Hamiltonian  is
\begin{equation}
\{\tau_\alpha,H+\pi_t\}_{p,q,\Phi,t,\pi_t}\approx 0,
\end{equation}
according to our previous Eq.(18).  Let us observe that  the solution of Eq.(24) in the extended phase space $ (q,p,\Phi,t,\pi_t)$ is exactly of  the same  form as the solution of Eq.(19),  after we consider the change $ \bar H\rightarrow H+\pi_t $. To this end we now define
\begin{equation}
H + \pi_t=\sum_{n=0}^{\infty}H^{(n)}, \qquad  H^{(n)}\sim \Phi^n.
\end{equation}
Since the term $\pi_t$ in the LHS contributes only to zero order in $\Phi$, the boundary condition
in  the above equation is
\begin{equation}
 H^{(0)}=  H_0(q,p,t) + \pi_t.
\end{equation}
In this way, the complete  expressions for the functions  $H^{(n)}, \ n\geq1,$ in Eq.(25) are given by the same equations  (22) and (23), where the bar is now removed to indicate the explicit time dependence of the Hamiltonian and  the constraints. The  Poisson brackets that had the subindex $(q,p)$ are now  calculated in the extended phase space  $(q,p,t,\pi_t)$.  Let us observe that the  explicit dependence of  the functions  $H^{(n)}$ upon $\pi_t$ appears only  in  $H^{(0)}$. Thus, the final Hamiltonian $H$ will be a function of $q,p,\Phi$ and $t$ only.  

The series expansion  procedure can be analogously  applied to calculate the corresponding extension of any observable $A_0(q,p,t)$  of our original problem.  Moreover, if we calculate  the Poisson brackets of  two of such augmented  functions and restrict  the result to $\Phi=0$,  we recover  the  Dirac bracket corresponding to  the original problem. In other words
\begin{eqnarray}
\nonumber
\{A,B\}_{(q,p,t,\pi_t,\Phi)}|_{\Phi=0} & = &
\{A^{(0)},B^{(0)}\}+\{A^{(1)},B^{(1)}\}_{(\Phi)}\\
&=&\{A^{(0)},B^{(0)}\}-\{A^{(0)},\phi_\alpha\}C^{\alpha\beta}\{\phi_\beta, B^{(0)}\}\\ \nonumber
&=&\{A_0,B_0\}^{*},
\end{eqnarray}
where  $\{,\}^*$ is the original  Dirac bracket, $A^{(0)}=A_0$ and $B^{(0)}=B_0$.  We remind the reader that all Poisson brackets without subindices are calculated in the original phase space $(q,p)$. Another interesting case is   when $ B=H$,  $(B_0=H_0+\pi_t )$,  which leads to 
\begin{eqnarray}
\nonumber
\frac{dA}{dt}|_{\Phi=0}&=&\{A,H\}_{(q,p,t,\pi_t,\Phi)} |_{\Phi=0}= \{A_0,H_0+\pi_t \}^{*}_{(q,p,t,\pi_t)}\\   
& = &\frac{\partial A_0}{\partial t}+\{A_0, H_0\}^*-\{A_0,\phi_\alpha\}C^{\alpha\beta}\frac{\partial\phi_\beta}{\partial t}, 
\end{eqnarray}
which is exactly the relation (4),  assuming  no first-class constraints present.

\section{BRST-BFV for the non-stationary case}

Since the Batalin-Tyutin's conversional approach
allows us to transform all second-class constraints into first-class
constraints, including the case when we have explicit  time dependence,  we can 
apply  the  BRST-BFV method of quantization to  arbitrary non-stationary systems. In this section we reformulate the results of \cite{baly} (and extend some results of \cite{h}) for the
non-stationary case and give an outline of the applicability of the FV theorem in this situation. For simplicity, we
consider only the bosonic case.

In order to prove that the BFV method is still applicable to time-dependent systems,  let us consider the reparametrized problem with the action defined by Eq. (7) but with first class constraints only (we assume  that the Batalin-Tyutin conversional  method of the previous section has been applied,  in such a way that  no  second class constraints remain in the problem).

Our method consists in   applying  the  standard BRST-BFV procedure to this canonical problem. We will subsequently  show that, after  choosing  a convenient fermionic  canonical gauge  and after performing  some  apropriate functional  integrals,   we are able to obtain the  expression of the BRST-BFV formulation for the original non-stationary  problem given by the  action (1),  with  first-class constraints only . From the point of view  of the calculation of the evolution operator, this procedure can be considered as the inverse of the reparametrization  program, whereby starting from the  action (7)  we obtain the propagator corresponding to the time-dependent action (1).

As we know from the discussion in section 2, reparametrization invariance introduces an additional first-class constraint $\psi$, in such a way that  our complete set of fist-class constraints is now 
\begin{equation}
\psi=\pi_t+H_0\approx 0,\quad \psi_A(q,p,t)\approx 0.
\end{equation}
From now on we introduce the index $a=(0,A)$, with the choice $\psi_0=\psi$ and we consider the phase space of canonical  variables $q,p,t,\pi_t  $, which evolve according to  the  parameter $s$. Following the standard steps,
 we promote  the Lagrange multipliers $\lambda^a$, $a=0,1,2,...m$ to dynamical
variables associating to each of them a
real momentum $\pi_a$, with the same Grassmann parity,  such  that
\begin{equation}
 \{\pi_a, \lambda^b\}=-\delta_a^b.
\end{equation}
These momenta are
constrained to vanish in order not to change the dynamical content of
the theory. 
Denoting  by $G_\alpha$, the $2m+2$ total number of  constraints (including $ \pi_0\approx0,  \ \pi_A\approx0$)   we define the vectors
\begin{equation}
G_\alpha=(\pi_a,\psi_a),\quad\eta^\alpha=(-i\P^a,C^a),\quad \P_\alpha=(i\bar C_a, \bar \P_a),
\end{equation}
with $G_0=(\pi_0,\psi_0\equiv\psi)$,  $G_\alpha\equiv(\pi_A,\psi_A)$ for $\alpha\ne 0$ and where $ (\P^a , \bar C_a  )$ together with $ (C^a, \bar \P_a)$ are canonically conjugated  odd Grassmann variables. 

Next we construct the canonical  BRST generator  $ \tilde\Omega$ in the extended phase-space $ (q,p,t,\pi_t, \lambda, \pi_\lambda, \eta, \P)$ following the standard prescription of Ref.\cite{h}. The effective action for this  system is 
\begin{equation}
S_{eff}=\int_{s_1}^{s_2}\Bigg({ dq^i \over ds}p_i+ { d t \over ds}\pi_t-\lambda^a {d \pi_a \over ds}+{ d \eta^\alpha \over ds}\P_\alpha-{\tilde H}_{eff}\Bigg)ds
\end{equation}
with
\begin{equation}
{\tilde H}_{eff}={\tilde H}_{BRST}-\{\tilde K , \ \tilde \Omega\},
\end{equation}
where  ${\tilde H}_{BRST} $ is  identically zero because of  reparametrization invariance.
The evolution operator is given by 
\begin{equation}
Z_{\tilde K}=\int{ \D\mu} \   exp(i S_{eff}),
\end{equation}
where $\D\mu $ is the Liouville measure in the above mentioned extended phase-space.

The basic idea in what follows is to make explicit the dependence of $S_{eff}$ upon the extra canonical  variables ( with respect to the original problem): $t, \ \pi_t, \lambda^0, \pi_0, C^0,   \P^0, {\bar C}_0, \bar\P_0   $ and to perform the corresponding functional integrals in the evolution operator (34), so that  we are  left with   the evolution operator associated  with  the original
non-stationary problem.  To this end, let us first split  $\tilde  \Omega$ in the form
\begin{equation}
\tilde \Omega= \eta^\alpha G_\alpha+\mbox{``more''}={\tilde \Omega }_{min} -i\P^0\pi_0 -i\P^A\pi_A,  
\end{equation}
where ${{\tilde \Omega }_{min}=\tilde \Omega }_{min}(t,\pi_t,q,p,C^0,\bar \P_0,C^A, \bar \P_A )$ is still nilpotent . By making explicit  the dependence  of ${\tilde \Omega }_{min}$ upon $ C^0$,   the new functions $h$ and $\Omega_{min}$  are defined, such that
\begin{equation}
 {\tilde \Omega }_{min} =C^0 h + \Omega_{min}.
\end{equation}

A direct  extension of the Theorem (6.1) of Ref.\cite{h}  allows us to prove that $h$ is linear in $\pi_t$, { \it  i.e.} $ h=(\pi_t + H_{BRST})$ and that both $H_{BRST} $ and $\Omega_{min} $ depend only upon the variables $t,q,p, C^A,   \bar \P_A$ and not  either on  $ \bar \P_0$ or on $ \pi_t$. Let us remind the reader  that our notation differs from that of  of Ref.\cite{h} : our original first-class constarints are labeled by the subindex $A$, while the extended constraints, which include $ \psi_0=\pi_t+ H_0$ are labeled by the subindex $a$.  
Such an extension is necessary because in our case $ \lambda =\pi_t$ has non-zero Poisson bracket with one of the remaining canonical variables : $t$.  Nevertheless, the proof of the above assertions follows the same steps given in Ref.\cite{h} for the time-independent case.  The reasons  are  basically the following  :  (i)  The zeroth-order  dependence upon  $\pi_t$ comes  only from $ \U{0}=\eta^0(\pi_t+H_0 (q,p,t))+\eta^AG_A(q,p,t)$. (ii)  The higher order  structure functions $ \U{n} $  can only  depend  on $\pi_t$ through the term $ \{ \U{n}, \ \U{0} \} $. This Poisson bracket   does not introduce an aditional  dependence on $ \pi_t$, but produces a further term proportional to $ {\partial \over \partial t}{\U{n}}$ which was not present  in the time-independent case. (iii) Another difference that appears in our case  is that the structure functions $C^A_{B0}= V^A_B $ are defined by \  $ [\psi_A, H_0]+\frac{\partial \psi_A}{\partial t}=V^C_A \psi_C$. (iv) Again, the structure functions $ \U{n}{}^{a_1\dots a_n}$ can be chosen to be zero whenever one of the upper indices $a_i$ is equal to zero. This is a direct consequence of the independence of the ghost and implies that $\tilde \Omega_{min}$ is independent of   $ \bar \P_0$ .

The nilpotency of  $ {\tilde \Omega }_{min} $,   together with the specific dependence upon the canonical variables of the  involved functions leads to the properties
\begin{equation}
\{\Omega, \ \Omega\}=0, 
\end{equation}
\begin{equation}
\{h, \ \Omega\}=0={\partial {\Omega}\over \partial t}+\{H_{BRST}, \ \Omega\},
\end{equation}
where we have reintroduced the full BRST charge  $ \Omega= \Omega_{min} -i\P^A\pi_A $. 
The above properties, together with the final expression that we will obtain for the evolution operator (34), once we have performed the extra functional integrals, lead to the interpretation of $ \Omega$  and $ H_{BRST}$ as the BRST-charge and BRST-Hamiltonian respectively,  corresponding to the original non-stationary problem. Once more we emphasize that both quantities are directly obtained from the construction of the canonical BRST-charge ${\tilde \Omega }_{min} $. 

Now we perform the extra functional integrals referred to above. With this purpose we choose
the canonical  fermionic gauge
\begin{equation}
\tilde K=\frac{i}{\epsilon}\bar C_0\chi^0-\bar\P_0\lambda^0+K^\prime.
\end{equation}
Here $K^\prime$ is an arbitrary  fermionic gauge which does not depend  either on the coordinates or the momenta with index 0 in the extended phase space, but which might depend on the  parameter $s$.  The function $ \chi^0$ will be subsequently selected  to  fix the gauge associated with reparametrizations. Substituting the above  choice of the fermionic gauge, together with Eqs. (35), (36),  in Eq. (33)  leads to
\begin{equation}  
H_{eff}=-\frac{1}{\epsilon}\chi^0\pi_0-\frac{i}{\epsilon}\bar C_0[\chi^0, h ]
C^0- \lambda^0 h -i\bar\P_0\P^0-[K^\prime,\Omega].
\end{equation}
Using   Eq.( 40) in the effective action, making the change of  integration variables  
\begin{equation}
\bar C_0\to \epsilon \bar C_0\qquad \pi_0\to\epsilon\pi_0,
\end{equation}
 with Jacobian one, and choosing the gauge fixing function
\begin{equation}
\chi^0=\frac{s_1-s}{S}T-t_1+t,
\end{equation}
with $T\equiv t_2-t_1, \  S\equiv s_2 - s_1$,we are able to  implement the canonical gauge $\chi^0=0$ and  to guarantee  that the correct end point conditions  are satisfied after taking the limit $\epsilon\to 0$ \cite{hd} .
Then,  the effective action can be rewritten as
\begin{eqnarray}
\nonumber
S_{eff}^{K^\prime}=\int_{s_1}^{s_2}&&ds \Bigg({d q^i \over ds}p_i+{ dt \over ds}\pi_t+{ d \P^A \over ds}\bar C_A + { d C^A \over ds}\bar\P_A
-\lambda^A { d \pi_A \over ds }\\
& +&  \{K^\prime,\Omega\}+{ d C^0 \over ds}\bar\P_0+\pi_0\chi^0+i \bar C_0\{\chi^0, h\}C^0+i\bar\P_0\P^0+\lambda^0 h \Bigg),
\end{eqnarray}
where $\{\chi^0,  h \}=1$.  After  performing  the functional  in\-te\-grals over  the ghosts $C^0,  \P_0,  \bar C_0,\bar\P_0$, and   $\lambda_0$ we obtain a delta functional that we use to integrate over $\pi_t$. Then, we integrate  $\pi_0$ to obtain a delta functional of the gauge condition, that  allows us to integrate  over $t $.  The effective action reduces to
\begin{equation}
{S_{eff}^{K^\prime}=\int_{s_1}^{s_2}ds \Bigg(\frac{dq^i}{ds}p_i+\frac{d\P^A}{ds}\bar C_A + \frac{d C^A}{ds}\bar\P_A-
\lambda^A\frac{d\pi_A}{ds}-\frac{T}{S}H_{BRST}+\{K^\prime,\Omega\}\Bigg),}
\end{equation}
with
\begin{equation}
t=\frac{s-s_1}{S}T+t_1. 
\end{equation}
If we now make a redefinition of the  fermionic gauge :  $K^\prime=\frac{T}{S}K$ and use Eq.(45) to write  $ds=\frac{S}{T}dt$,  we obtain the following  effective action for the original problem
\begin{equation}
{S_{eff}^K=\int_{t_1}^{t_2}dt \Bigg(\frac{dq^i}{dt}p_i +\frac{d\P^A}{dt}\bar C_A + \frac{d C^A}{dt}\bar\P_A-\lambda^A \frac{d\pi_A}{dt} -H_{BRST}(q,p,t)+\{K,\Omega\}
\Bigg)}
\end{equation}

This is the main result of this paper, which amounts to the calculation of the effective action to be used in the application of the BRST-BFV method for the time-dependent case. The above expression (46) allows for the discussion of the applicability of the FV theorem to this situation. Following   steps analogous  to those of  the proof given in theorem 9.1 of  Ref.\cite{h} for the time-independent case,  we are able to verify  that $S_{eff}$ is indeed independent of the fermionic gauge $K$. 
 Let us observe that expression (46) is what one would have naively expected, except maybe for the   condition (38) upon $H_{BRST}$ which enforces the statement that $\Omega$ is a conserved  time-dependent charge. Since  $H_{BRST}$ is explicitly time dependent,  the result that it is not a BRST-observable seems rather reasonable. It is worth noticing that Eq. (38), which     defines $H_{BRST}$ once $\Omega$ is constructed according to Eq.(37),  is invariant under the change  $ H_{BRST} \rightarrow  H_{BRST}  + [ K, \ \Omega]$ independently of the fact that $H_{BRST}$ is not an observable.   

Before closing this section we make some comments  that  will permit the direct calculation
of $ \Omega$ and $ H_{BRST}$ to be used in Eq.(46), without having to go through the reparametrized construction in each case.
From  Eq.  (37),  together with the properties of the functions involved, we conclude that the construction of the structure
functions, leading  to  the  nilpotent BRST charge $\Omega$,  is unchanged with respect to
the stationary case, except for the explicit time dependence that arises now. 

Nevertheless,  Eq. (38) implies that some modifications appear  when considering the construction of   $ H_{BRST}$ in  the time-dependent case. To make contact with the standard construction let us recall that Eq.(38) can be rewritten as
\begin{equation}
\{\Omega , \  H_{BRST} + \pi_t   \}_{(\dots ,t, \pi_t)}=0, 
\end{equation}
which  has  exactly the same form as  the   corresponding equation  that satisfies the
BRST-Hamiltonian  in the time independent case, except for the $ (t, \pi_t)$ extension of the phase space.  Since we know the
solution to the former  case \cite{h}, we can apply  it   directly to the non-stationary situation by defining
\begin{equation}
{ H}_{BRST} + \pi_t=\sum_{n\ge 0}{H}^{(n)},
\end{equation}
 where the only difference is again  the boundary condition 
\begin{equation}
{{ H}}^{(0)}=H_0 +\pi_t.
\end{equation}
In this sense, the situation is similar to that of the extension of the conversional approach decribed in Section 3. In the present  case however,  the series expansion (48) is  performed in terms of the ghost and anti-ghost fields in such a way that each  ${H}^{(n)} $ has ghost-number zero. The detailed  recurrence relations that allow for  the construction of the functions $ {H}^{(n)}$ can be found in Ref.\cite{h}.

\section{The BRST-BFV quantization of the time-dependent two-dimensional rotor} 
 
In order to illustrate in a transparent way the general properties previously  described,   we discuss in   this section the example  of  the two-dimensional   rotor with  time dependent 
radius.  Following the comments at the end of Section 4, we have chosen to deal with this problem using directly the final results (34), (37), (38), and (46) obtained in the  previous section,  instead of  starting again from a parametrized version of the system. 

The Lagrangian of the system can be written as
\begin{equation}
L=\frac{m}{2}({\dot r}^2+r^2{\dot \theta}^2)-\lambda (r-a(t)),
\end{equation}
where $m$ is the mass of the particle, $r,\theta$ are  polar coordinates in 
the plane,  $a(t)$  is the time dependent radius and $ \lambda$ is a Lagrange multiplier. The associated momenta are
$
p_r=m{\dot r}$ and  $ p_\theta=mr^2{\dot\theta},
$ which lead to  the  total Hamiltonian
\begin{equation}
H_{T0} =H_0 + \lambda (r-a) = {\frac{1}{2m}}\left( p_r^2 + {\frac{\po^2 }{r^2 }}%
\right) +\lambda (r-a),
\end{equation} 
together with the  constraints 
\begin{equation}
\xi_1=r-a(t)\approx 0, \qquad \xi_2=p_r-m\dot a(t)\approx 0.
\end{equation} 
The compatibility condition  ${d{\xi_2} \over dt} \approx 0$, allows us to 
determine  the Lagrange multiplier $\lambda$ as
\begin{equation}
\lambda=\frac{p_\theta^2}{mr^3}-m{\ddot a}.
\end{equation}
In this way the Dirac procedure stops and we conclude that the constraints (52)  are second-class and time-dependent . 

In order to carry on the BRST-BFV quantization procedure it is necessary to 
start from a system including only first-class constraints. To this end we 
apply the conversional  procedure described in section 3,  to the  second class constraints $\xi _1,\xi _2$.  This requires the addition of  two  canonically conjugated  variables $q$ and $\pi  
$, such that $\left\{ q,\pi \right\} =1$ . A direct application of the method leads to  the augmented first class constraints 
\begin{equation}
\tau_1=\xi_1+q, \qquad \tau_2=\xi_2-\pi.
\end{equation}
The  total augmented   Hamiltonian $H_T$ can be constructed  according to Eqs. (25), (26), (22) and (23). Starting from  
\begin{equation}
H^{(0)}= H_{T0} + \pi_t=\frac{1}{2m}(p_r^2+\frac{p_\theta^2}{r^2})+\lambda(r-a)+\pi_t,
\end{equation}
we calculate the first three terms in the expansion (25)  
\begin{eqnarray}
\nonumber
H^{(1)}&=&-q(\frac{p_\theta^2}{mr^3}-\lambda-m\ddot a) - \pi(\frac{p_r}{m}-\dot a),\\
H^{(2)}&=&\frac{1}{2m}(3q^2\frac{p_\theta^2}{r^4}+\pi^2),\\ \nonumber
H^{(3)}&=&-\frac{6q^3}{4m}\frac{p_\theta^2}{r^5}.
\end{eqnarray}
From these relations we can infer the form of the full augmented Hamiltonian if we make use of the binomial series for $(r+q)^{-2}$. The result is
\begin{equation}
H_T=\frac{1}{2m}[\pi_\eta^2+\frac{p_\theta^2}{Q^2}] +\lambda(Q-a(t))+m{\ddot 
a}(Q/2-\eta)+{\dot a}(\pi_Q-\pi_\eta/2),
\end{equation}
where we have introduced  new  canonical  variables
\begin{equation}
\eta=\frac{1}{2}(r-q),\quad \pi_\eta=p_r-\pi, \quad 
Q=r+q,\quad \pi_Q=\frac{1}{2}(p_r+\pi).
\end{equation}
We have explicitly verified that the above expression (57) does preserve in time  the first-class  constraints (54). The full  extended Hamiltonian is $ {\cal H}_E={ \cal H}_0+\lambda ^\alpha \tau_\alpha $
where ${\cal H}_0$ is defined by subtracting the term $\lambda {\tau }_1$ in Eq. 
(57). The corresponding augmented first order Lagrangian function is 
\begin{equation}
L_E=-{\dot \pi}_Q Q + p_\theta{\dot 
\theta}+\pi_\eta{\dot 
\eta}-\frac{1}{2m}\left[\pi_\eta^2+\frac{p_\theta^2}{Q^2}\right] 
-m{\ddot a}(Q/2-\eta)$$ $$-{\dot 
a}(\pi_Q-\pi_\eta/2)-\lambda(Q-a)-\sigma(\pi_\eta-m{\dot a}). 
\end{equation}
 
Following the standard BRST formulation, we define the vectors
\begin{eqnarray}
\nonumber
G_\alpha&=&(\pi_\lambda,\pi_\sigma,Q-a,\pi_\eta-m\dot a),\\ 
 \eta^\alpha&=&(-i\P^1,-i\P^2,C^1,C^2),\\ \nonumber \P_\alpha&=&(i{\bar C}_1,i{\bar 
C}_2,{\bar \P}_1,{\bar \P}_2),
\end{eqnarray}
where the last two correspond to the ghosts and anti-ghosts respectively, 
which have the Poisson brackets $\{\eta^\alpha,\P_\beta\}=\{\P_\beta,\eta^\alpha\}=-{\delta^\alpha_\beta}$. 
The BRST charge $\Omega $ is given by
\begin{equation}
\Omega=-i\P^1\pi_\lambda-i\P^2\pi_\sigma+C^1(Q-a)+C^2(\pi_\eta-m\dot a).
\end{equation}
The evolution operator is determined by the effective quantum action
$
Z = \int \D \mu \exp(iS_{eff} ),
$
where $\D \mu $ is the Liouville measure corresponding to all canonical 
variables involved.  According to our result  (46) the expression for $S_{eff} $ is 
\begin{equation}
S_{eff}= \int^{t_2}_{t_1} dt \left(\dot q^i p_i -\lambda^a \dot \pi_a + \dot 
\eta^\alpha\P_\alpha- H_{eff} \right).
\end{equation}
The effective Hamiltonian is defined by  $H_{eff} = H_{BRST} - \left \{ K  , \Omega \right\},$
where $K $ is  the fermionic gauge fixing term. Since the extended 
Hamiltonian  $H_E$  already satisfies Eq.(38), we take $H_{BRST}=H_E$. Imposing the fermionic  
gauge condition
$
K=\bar{\P_1}\lambda+\bar{\P_2}\sigma,
$
the effective Hamiltonian can be written as 
\begin{eqnarray}
\nonumber H_{eff}&=&\frac{1}{2m} 
\left(\pi_\eta^2+\frac{p_\theta^2}{Q^2}\right)+ i\bar{\P_1}\P^1 
+i\bar{\P_2}\P^2 \\ 
&+&\lambda(Q- a)+\sigma(\pi_\eta-m\dot a)+m\ddot 
a(\frac{Q}{2}-\eta)+\dot a(\pi_Q-\frac{\pi_\eta}{2})
\end{eqnarray}
and the classical effective action reads
\begin{eqnarray}
\nonumber
S_{eff}&=&\int_{t_1}^{t_2}dt 
\Big[p_\theta\dot\theta+\pi_\eta\dot\eta-\dot\pi_QQ- 
{\dot\pi}_\lambda\lambda- {\dot\pi}_\sigma\sigma+ {\dot\P}^1{\bar 
C}_1+{\dot\P}^2{\bar C}_2+{\dot C}^1{\bar\P}_1\\ & +& {\dot 
C}^2{\bar\P}_2-\frac{p_\theta^2}{2mQ^2}- 
\frac{\pi_\eta^2}{2m}-m{\ddot a}(\frac{Q}{2}-\eta) -{\dot 
a}(\pi_Q-\frac{\pi_\eta}{2})-i{\bar\P}_1{\P}^1\\ 
\nonumber
& - & i{\bar\P}_2{\P}^2-\lambda(Q-a)- \sigma(\pi_\eta-m{\dot a})\Big].
\end{eqnarray}

Now we consider the appropriate end-point conditions for the corresponding 
variables. These conditions must be BRST-invariant and they should provide a 
unique solution for the equations of motion derived from the above effective 
action. 
 
In our case, the ghosts and anti-ghosts are not coupled to the remaining variables 
and we obtain  the equations
\begin{equation}
\ddot{{\bar C}_{k}}=0,\quad\ddot{{C}^{k}}=0.
\end{equation}
Then,  it is enough to choose the following end-point conditions for this 
sector of the problem
\begin{eqnarray}
\nonumber C^1(t_1)&=&C^2(t_1)=C^1(t_2)=C^2(t_2)=0,\\ {\bar 
C}^1(t_1)&=&{\bar C}^2(t_1)={\bar C}^1(t_2)={\bar C}^2(t_2)=0.
\end{eqnarray}
  
The remaining equations of motion are
\begin{eqnarray}
\nonumber {\dot p_\theta=0}&\qquad& 
{\dot\theta}-\frac{p_\theta}{mQ^2}=0,\\ \nonumber {\dot\pi_\eta}-m{\ddot 
a}=0&\qquad& {\dot\eta}-{\pi_\eta/m}+{\dot a}/2-\sigma=0,\\ 
{\dot\pi_Q}-\frac{p_\theta^2}{mQ^3}+\frac{m\ddot 
a}{2}+\lambda=0&\qquad& {\dot Q}-{\dot a}=0, \\
\nonumber \dot \lambda=0&\qquad& 
{\dot \pi}_{\lambda} +Q-a=0,\\ \nonumber \dot\sigma=0&\qquad &{\dot \pi}_{\sigma} 
+{\pi}_{\eta}-m\dot a=0.
\end{eqnarray} 
In particular,   they imply ${\ddot{\pi}}_{\lambda}=0= {\ddot{\pi}}_{\sigma}$ 
so that we can impose the following boundary conditions
\begin{equation}
\pi_\lambda(\tau_1)=\pi_\lambda(\tau_2)=\pi_\sigma(\tau_1)= 
\pi_\sigma(\tau_2)=0,
\end{equation}
in order to guarantee that ${\pi}_{\lambda}(t)=0={\pi}_{\sigma}(t) $. In 
this way we recover the original constraints $Q-a=0$ and ${\pi}_{\eta}-m\dot 
a=0 $ as a consequence of the equations of motion. The remaining 
second-order differential equations that decouple the  system of 
equations (67) are
\begin{equation}
{\ddot{\eta}- {1\over 2}\ddot{a} =0}, \qquad 
\ddot{\theta}+{2{p}_{\theta}\over ma^3} \dot \theta=0, \qquad  
{\ddot{\pi}}_{Q} + {3{{p}_{\theta}}^2\over ma^4} \dot 
a+m\ddot{a}=0.
\end{equation} 
Unique solutions to the above equations are obtained by fixing the 
end-points of the corresponding variables. Denoting by $z(t_1)=z_1, \ 
z(t_2)=z_2$ the corresponding fixed values, the remaining variables ${p}%
_{\theta}, \sigma, \lambda$ are uniquely determined in the following form
\begin{eqnarray}
\nonumber p_\theta=\frac{m(\theta_2-\theta_1)}{\int_{t_1}^{t_2}a^{-2}dt}, 
\qquad 
\sigma=\frac{\eta_2-\eta_1}{t_2-t_1}-\frac{a_2-a_1}{2(t_2-t_1)}, 
&&\\
\quad 
\lambda=\frac{p_\theta^2}{m(t_2-t_1)}\int_{t_1}^{t_2}a^{-3}dt-\frac{m({\dot 
a}_2-{\dot a}_1)}{t_2-t_1}-\frac{{\pi_Q}_2 - 
{\pi_Q}_1}{t_2-t_1}.
\end{eqnarray}
Now we further specify the above boundary conditions. We choose
\begin{equation}
\eta_1= {\frac{1}{2}}a({t}_{1})\qquad\eta_2= {\frac{1}{2}}a({t}_{2}), 
\end{equation}
which leads to $q(t)=0$ implying also that $\sigma=0$. Our next choice is
\begin{equation}
\pi_{Q_1}=\pi_{Q_2}=0,
\end{equation}
in such a way that Eq.(70) leads to the correct Lagrange multiplier $\lambda$ 
in the time-independent case. We have verified that all the imposed boundary 
conditions are in fact BRST-invariant. 
 
Next we calculate the integral 
\begin{eqnarray}
\nonumber Z&=&\int\D\P^1\D{\bar\P}_1\D C^1\D{\bar 
C}_1\D\P^2\D{\bar\P}_2\D C^2\D{\bar C}_2\\ 
&&\D\pi_\eta\D\eta\D 
p_\theta\D\theta\D\pi_Q\D Q \D\pi_\lambda\D 
\lambda\D\pi_\sigma\D\sigma \exp[iS_{eff}].
\end{eqnarray}
 
The integration over the ghosts gives an overall factor of $T^2$ where $%
T=\tau_2-\tau_1$, so that $Z $ is reduced to
\begin{eqnarray}
\nonumber Z&=&T^2\int\D\pi_\eta\D\eta\D 
p_\theta\D\theta\D\pi_Q\D Q \D\pi_\lambda\D 
\lambda\D\pi_\sigma\D\sigma \\ 
&&\exp\Big[i\int^{\tau_2}_{\tau_1}d\tau 
\big(p_\theta\dot\theta+ 
\pi_\eta\dot\eta-\dot\pi_QQ- 
{\dot\pi}_\lambda\lambda-{\dot\pi}_\sigma\sigma -\frac{p_\theta^2}{2mQ^2}- 
\frac{\pi_\eta^2}{2m}\\ \nonumber 
&-&m{\ddot a}(\frac{Q}{2}-\eta)- 
{\dot a}(\pi_Q-\frac{\pi_\eta}{2}) 
-\lambda(Q-a)-\sigma(\pi_\eta-m{\dot a})\big) \Big]
\end{eqnarray}
   
The remaining functional integrations are calculated by using the general 
expression
\begin{eqnarray}
\nonumber
\int \D q &&\D p \exp i \int_{{t}_{1}}^{{t}_{2}}{dt (p\dot q+ F(p;z) +q\dot 
g(t))}= 
\\ 
\int_{- \infty}^{+ \infty}{d{p}_{0}}&&\exp i \left\{ {p}_{0}({q}_{1}-{q}_{2}) +  
{q}_{1}(g({t}_{2})-g({t}_{1})) \right\} \times\\ \nonumber &&\exp i \int_{t_1}^{t_2}{dt 
F(p_0+g(t)-g(t_2); z)},
\end{eqnarray}
where the variable $p$ is free at the end-points, while $q$ is fixed by the 
conditions $q(t_1)=q_1, \ q(t_2)=q_2$. We have denoted by $z$ any other 
variable involved and $g(t) $ is an arbitrary function of time . 
 
The integration over $\sigma, \pi_\sigma, \lambda, \pi_\lambda$ correspond 
to the above case with $g=0$ and the result is 
\begin{eqnarray}
\nonumber Z&=& T^2 \int d\sigma_0 d\lambda_0\D\pi_\eta\D\eta\D 
p_\theta\D\theta\D\pi_Q\D Q\\ &&\exp\Big[i\int^{\tau_2}_{\tau_1}d\tau 
\big(p_\theta\dot\theta+ \pi_\eta\dot\eta-\dot\pi_QQ 
-\frac{p_\theta^2}{2mQ^2}- \frac{\pi_\eta^2}{2m}\\ \nonumber
&-&m{\ddot 
a}(\frac{Q}{2}-\eta)- {\dot a}(\pi_Q-\frac{\pi_\eta}{2}) 
-\lambda_0(Q-a)-\sigma_0(\pi_\eta-m{\dot a})\big) \Big]
\end{eqnarray}

The remaining functional integrations over $\eta$, $\pi_\eta$ and $\pi_Q$, $Q 
$ can also be done according to the above formula . The integrations with 
respect to $\lambda_0$ and $\sigma_0$ contribute each with a factor of $1/T$ 
and produce adequate $\delta$-functions. The final result is
\begin{equation}
Z =\int \D 
p_\theta \D\theta \exp \Big[i\int_{\tau _1}^{\tau _2}d\tau \big( p_\theta {%
\dot \theta }-\frac{p_\theta ^2}{2ma^2(\tau )}-\frac m2{\dot a}^2(\tau )\big)%
\Big].
\end{equation}
The last term in the resulting expression for $Z $ comes from the term $%
p_r^2/(2m)$ from the original variational principle. This term itself  is  
a total time derivative  and contribute to $Z $ only with an overall 
phase factor. The resulting expression for $Z $,  up to a phase factor, is \cite{k} 
\begin{equation}
Z =\langle \theta _2\tau _2|\theta _1\tau _1\rangle =\sum_{n=-\infty 
}^\infty \frac 1{2\pi }\exp \left[ in(\theta _2-\theta _1)-i\frac{n^2}{2m}%
\int_{\tau _1}^{\tau _2}d\tau \frac 1{a^2(\tau )}\right]
\end{equation}
In this form we recover the well known expression for the propagator 
of the time dependent rigid rotor.  
 
\section{Acknowledgements}

  LFU, JDV and JAG were partially supported by the grant UNAM-DAGAPA-IN100694. LFU and JDV also ac\-knowledge support fron the grant CO\-NA\-CyT-400200-5-3544E. JAG is supported by the  CONACyT  graduate fellowship \# 86226. LFU  was also partially supported by a fellowship from Fundaci\'on Andes, Chile and by the  grant CONACYT(M\'exico)-CONICYT(Chile) \# E120.2778.  He   acknowledges the kind hospitality of  J.   Alfaro at  Pontificia Universidad Cat\'olica de Chile.  He also
 thanks  M. Henneaux and C. Teitelboim for useful suggestions.

\end{document}